\documentclass[preprint,aps]{revtex4}

\usepackage{graphicx}
\usepackage{dcolumn}
\usepackage{bm}

\usepackage{amssymb}
\usepackage{latexsym}
\usepackage{amsthm}
\usepackage{graphicx}
\theoremstyle{plain}

\newtheorem{thm}{Theorem}

\newtheorem{cor}{Corollary}
\newtheorem{pro}{Proposition}

\newcommand{\ZM}{\bf Z}

\newcommand{\CM}{\bf C}

\begin{document}

\title{Limit Theorem for Continuous-Time Quantum Walk \\ on the Line}

\author{Norio Konno}
\email{norio@mathlab.sci.ynu.ac.jp}
\affiliation{%
Department of Applied Mathematics, 
Yokohama National University, 
79-5 Tokiwadai, Yokohama, 240-8501, Japan\\}

\date{\today}

\vskip 2.0cm

\clearpage

\begin{abstract}
Concerning a {\it discrete-time} quantum walk $X^{(d)} _t$ with a symmetric distribution on the line, whose evolution is described by the Hadamard transformation, it was proved by the author that the following weak limit theorem holds: $X^{(d)} _t /t  \to dx / \pi (1-x^2) \sqrt{1 - 2 x^2}$ as $t \to \infty.$ The present paper shows that a similar type of weak limit theorems is satisfied for a {\it continuous-time} quantum walk $X^{(c)} _t$ on the line as follows: $X^{(c)}_t /t  \to dx / \pi \sqrt{1 - x^2}$ as $t \to \infty.$ These results for quantum walks form a striking contrast to the central limit theorem for symmetric discrete- and continuous-time classical random walks: $Y_{t}/ \sqrt{t} \to e^{-x^2/2} dx / \sqrt{2 \pi}$ as $t \to \infty.$ The work deals also with issue of the relationship between discrete and continuous-time quantum walks. This topic, subject of a long debate in the previous literature, is treated within the formalism of matrix representation and the limit distributions are exhaustively compared in 
the two cases.
\end{abstract}

\pacs{02.50.Cw, 03.67.Lx, 05.40.Fb}
\maketitle

\clearpage

\section{\label{sec:S1} Introduction}
Quantum walks have recently been introduced and investigated, with the hope that they may be useful in constructing new efficient quantum algorithms, (for reviews of quantum walks, see \cite{Amb,Kem,TFMK}). There are two distinct types of the quantum walk: one is a discrete-time case \cite{AAKV,ABNVW,GJS,HKS,Kon2002,Kon2005,Mey,PRR,RSSAAD}, the other is a continuous-time case \cite{AAHT,ABTW,CCDFGS,CFG,FG,IKKK,MR}. The quantum walk can be considered as a quantum analog of the classical random walk. However there are some differences between them. For the discrete-time symmetric {\it classical} random walk $Y^o _t$ starting from the origin, the central limit theorem shows that $Y^o _{t}/ \sqrt{t} \to e^{-x^2/2} dx / \sqrt{2 \pi}$ as $t \to \infty.$ The same limit theorm holds for the continuous-time classical symmetric random walk, (related results can be found in Refs. \cite{Dur,GS}). The limit density function is the normal distribution and has a bell-shaped curve with one peak at the center. On the other hand, concerning a {\it discrete-time} quantum walk $X^{(d)} _t$ with a symmetric distribution on the line, whose evolution is described by the Hadamard transformation, it was shown by the author \cite{Kon2002,Kon2005} that the following weak limit theorem holds: $X^{(d)} _t /t  \to dx / \pi (1-x^2) \sqrt{1 - 2 x^2}$ as $t \to \infty.$ This paper presents that a similar type of weak limit theorem is proved for a {\it continuous-time} quantum walk $X^{(c)} _t$ on the line as follows: $X^{(c)}_t /t  \to dx / \pi \sqrt{1 - x^2}$ as $t \to \infty.$ Both limit density functions for quantum walks have two peaks at the two end points of the support. The study of weak limits for discrete-time quantum walks is treated also in Ref. \cite{GJS}, with a simplified proof with respect to earlier derivations. 

As a corollary, we have the following result. Let $\sigma^c _{(d)} (t)$ (resp. $\sigma^c _{(c)} (t)$) be the standard deviation of the probability distribution for a discrete-time (resp. continuous-time) classical random walk on the line at time $t$. Similarly, $\sigma^q _{(d)} (t)$ (resp. $\sigma^q _{(c)} (t)$) denotes the standard deviation for a discrete-time (resp. continuous-time) quantum walk. Then, the central limit theorem implies that $\sigma^c _{(d)} (t), \> \sigma^c _{(c)} (t) \asymp \sqrt{t},$ where $f(t) \asymp g(t)$ indicates that $f(t)/g(t) \to c_{\ast} (\not= 0)$ as $t \to \infty.$ In contrast, using our limit theorems, it is shown that $\> \sigma^q _{(d)} (t), \> \sigma^q _{(c)} (t) \asymp t$ hold. That is, the qunatum walks spread over the line faster than the classical walks in the both discrete- and continuous-time cases. 

The remainder of this paper is organized as follows. Sect. $\mbox{I\hspace{-0.5mm}I}$ gives the definition of the walk and our results. In Sect. $\mbox{I\hspace{-0.5mm}I\hspace{-0.5mm}I}$, we prove Proposition 1. Sect. $\mbox{I\hspace{-0.5mm}V}$ is devoted to a proof of Theorem 1. Conclusion and discussion are given in Sect. $\mbox{V}$.

\section{Model and Results}
Let $\ZM$ be the set of integers. To define the continuous-time qunatum walk on $\ZM$, we introduce an $\infty \times \infty$ adjacency matrix of $\ZM$ denoted by $A$ as follows: 
\begin{eqnarray*}
A \> = \> 
\bordermatrix{
   & \ldots & -3 & -2 & -1 & \>\>\>0 & +1 & +2 &\ldots \cr
\vdots & \ddots & \cdot  &  \cdot &  \cdot &  \cdot & \cdot & \cdot & \ldots \cr
-3 & \ldots &  0 &  1 &  0 & 0 &  0 &  0  & \ldots \cr
-2 & \ldots &  1 &  0 &  1 & 0 &  0 &  0  & \ldots \cr
-1 & \ldots &  0 &  1 &  0 & 1 &  0 &  0  & \ldots \cr
\>\>\> 0  &  \ldots & 0 &  0 &  1 & 0 & 1 & 0 & \ldots \cr
+1  &  \ldots & 0 &  0 &  0 & 1 & 0 & 1 & \ldots \cr
+2  &  \ldots & 0 &  0 &  0 & 0 & 1 & 0 & \ldots \cr
\vdots & \ldots & \cdot  &  \cdot &  \cdot &  \cdot & \cdot & \cdot & \ddots 
}.
\label{eqn:gashin}
\end{eqnarray*}
The amplitude wave function of the walk at time $t$, $| \Psi (t) \rangle$, is defined by
\begin{eqnarray*}
| \Psi (t) \rangle = U (t) | \Psi (0) \rangle ,
\label{eqn:evolution}
\end{eqnarray*}
where
\begin{eqnarray*}
U(t) = e^{itA/2} .
\label{eqn:unitary}
\end{eqnarray*}
Note that $U(t)$ is a unitary matrix. As an initial state, we take 
\begin{eqnarray*}
| \Psi (0) \rangle = {}^T (\ldots, 0, 0, 0, 1, 0, 0, 0, \ldots),
\label{eqn:hajime}
\end{eqnarray*}
where $T$ indicates the transposed operator. Concerning the details of the definitions for the continuous-time case, see \cite{AAHT,ABTW,GW}. Let $| \Psi (k,t) \rangle$ be an amplitude wave function at location $k$ at time $t.$ The probability that the particle is at location $k$ at time $t$, $P (k,t)$, is given by
\begin{eqnarray*}
P (k,t) = \langle \Psi (k,t) | \Psi (k,t) \rangle .
\label{eqn:evolution}
\end{eqnarray*}
As for results on the walk on a circle, i.e., $\{0,1, \ldots, N-1\}$, see \cite{AAHT,ABTW,IKKK}.

We will obtain an explicit form of $U(t)$ first. Our approach is a direct computation based on the $\infty \times \infty$ matrix $A$ without using its eigenvalues and eigenvectors in order to clarify a relation (stated below) between the continuous-time and discrete-time quantum walks, (another approach and the same explicit form can be found in Section III C of Ref. \cite{CCDFGS}). Let $J_k (x)$ denote the Bessel function of the first kind of order $k.$ As for the Bessel function, see Watson \cite{Wat} and Chapter 4 in Andrews {\it et al.} \cite{AAR}.
\begin{pro}
In our setting, we have
\begin{eqnarray*}
U (t) \> = \> 
\bordermatrix{
   & \ldots & -3 & -2 & -1 & \>\>\>0 & +1 & +2 & \ldots \cr
\vdots & \ddots & \cdot  &  \cdot &  \cdot &  \cdot & \cdot & \cdot & \ldots \cr
-3 & \ldots &  J_0 (t) &  i J_1 (t) &  i^2 J_2(t) & i^3 J_3 (t) & i^4 J_4 (t) &  i^5 J_5 (t)  & \ldots \cr
-2 & \ldots &  i J_1 (t) & J_0 (t)  &  i J_1 (t) & i^2 J_2 (t) &  i^3 J_3 (t) &  i^4 J_4 (t)  & \ldots \cr
-1 & \ldots &  i^2 J_2 (t) &  i J_1 (t) &  J_0 (t) & i J_1 (t) &  i^2 J_2 (t) &  i^3 J_3 (t) & \ldots \cr
\>\>\> 0  &  \ldots & i^3 J_3 (t) &  i^2 J_2 (t) &  i J_1 (t) & J_0 (t) & i J_1 (t) & i^2 J_2 (t) & \ldots \cr
+1  &  \ldots & i^4 J_4 (t) &  i^3 J_3 (t) &  i^2 J_2 (t) & i J_1 (t) & J_0 (t) & i J_1 (t) & \ldots \cr
+2  &  \ldots & i^5 J_5 (t) &  i^4 J_4 (t) &  i^3 J_3 (t) & i^2 J_2 (t) & i J_1 (t) & J_0 (t) & \ldots \cr
\vdots & \ldots & \cdot  &  \cdot &  \cdot & \cdot &  \cdot & \cdot & \ddots 
}.
\label{eqn:shotan}
\end{eqnarray*}
That is, the $(l,m)$ component of $U(t)$ is given by $i^{|l-m|} J_{|l-m|}(t).$
\end{pro}
From Proposition 1, looking at a column of $U(t)$, we have immediately 
\begin{cor}
The amplitude wave function of our model is given by  
\begin{eqnarray*}
| \Psi (t) \rangle = {}^T (\ldots, i^3 J_3 (t), i^2 J_2 (t),i J_1 (t), J_0 (t),  i J_1 (t), i^2 J_2 (t), i^3 J_3 (t), \ldots),
\label{eqn:amplitude}
\end{eqnarray*}
that is, $| \Psi (k,t) \rangle = i^{|k|} J_{|k|} (t)$ for any location $k \in \ZM$ and time $t \ge 0.$ 
\end{cor}
Moreover, noting that $J_{-k} (t) = (-1)^k J_k (t)$ (Eq. (4.5.4) in Ref. \cite{AAR}), we have
\begin{cor}
The prbability distribution is  
\begin{eqnarray*}
P(k,t) = J_{|k|} ^2 (t) = J_{k} ^2 (t),
\label{eqn:probmeas}
\end{eqnarray*}
for any location $k \in \ZM$ and time $t \ge 0.$
\end{cor}
In fact, the following result (see Eq. (4.9.5) in Ref. \cite{AAR}):
\begin{eqnarray*}
 J_0 ^2 (t) + 2 \sum_{k=1} ^{\infty} J_{k} ^2 (t) =1
\label{eqn:probmeasone}
\end{eqnarray*}
ensures that $\sum_{k=- \infty} ^{\infty} P(k,t) =1$ for any $t \ge 0.$ Remark that the distribution is symmetric for any time, i.e., $P(k,t)=P(-k,t).$

One of the interesting points of the above mentioned results is as follows. Let $\CM$ be the set of complex numbers. Fix a positive integer $r$. We suppose that a unitary matrix $U_r$ has the following form: 
\begin{eqnarray*}
U_r \> = \> 
\bordermatrix{
   & \ldots & -3 & -2 & -1 & \>\>\>0 & +1 & +2 & \ldots \cr
\vdots & \ddots & \cdot  &  \cdot &  \cdot &  \cdot & \cdot & \cdot & \ldots \cr
-3 & \ldots &  w_0  &  w_1 &  w_2 & w_3 & w_4  &  w_5  & \ldots \cr
-2 & \ldots &  w_{-1}  & w_0   &  w_1  & w_2  &  w_3  &  w_4   & \ldots \cr
-1 & \ldots &  w_{-2}  & w_{-1} & w_0   &  w_1  & w_2  &  w_3  & \ldots \cr
\>\>\> 0  &  \ldots & w_{-3} & w_{-2}  & w_{-1} & w_0   &  w_1  & w_2  & \ldots \cr
+1  &  \ldots & w_{-4} & w_{-3} & w_{-2}  & w_{-1} & w_0   &  w_1  & \ldots \cr
+2  &  \ldots & w_{-5} & w_{-4} & w_{-3} & w_{-2}  & w_{-1} & w_0  & \ldots \cr
\vdots & \ldots & \cdot  &  \cdot &  \cdot & \cdot &  \cdot & \cdot & \ddots 
}.
\label{eqn:nogo}
\end{eqnarray*}
with $w_k \in \CM $ ($k \in \ZM $) and $w_s = 0$ for $|s| > r$. Then {\it No-Go Lemma} (see Ref. \cite{Mey}) shows that there is the only non-zero $w_{\ast}$ with $|w_{\ast}|=1$; that is, there exists no non-trivial, homogeneous finite-range model governed by the $U_r$. For example, when $r=1$, we have ``$|w_{-1}|=1, w_0 = w_1= 0$'', ``$|w_{0}|=1, w_{-1} = w_1= 0$'', or ``$|w_{1}|=1, w_0 = w_{-1}= 0$''. Thus, the model has a trivial probability distribution. However, our homogeneous, but {\it infinite-range} model has a non-trivial probability distribution given by squared Bessel functions (Corollary 2).

An open problem on quantum walks is to clarify a relation between discrete-time and continuous-time quantum walks (see Ref. \cite{Amb}, for example). Another interesting point of the results is to shed a light on the problem. To explain the reason, we introduce the following matrices as in our previous paper \cite{HKS} :
\begin{eqnarray*}
&&
P_A= 
\left(
\begin{array}{cc}
a & b \\
0 & 0 
\end{array}
\right), 
\quad
Q_A=
\left(
\begin{array}{cc}
0 & 0 \\
c & d 
\end{array}
\right)
\quad \hbox{and} \quad
P_B= 
\left(
\begin{array}{cc}
a & 0 \\
c & 0 
\end{array}
\right), 
\quad
Q_B=
\left(
\begin{array}{cc}
0 & b \\
0 & d 
\end{array}
\right),
\end{eqnarray*}
where we assume that $U=P_j+Q_j \> (j = A, B)$ is a $2 \times 2$ unitary matrix. Here we consider two types of the discrete-time case; one is A-type, the other is B-type. The precise definition is given in Ref. \cite{HKS}. Then the unitary matrix of the discrete-time quantum walk on the line is described as 
\begin{eqnarray*}
U_{(d)} = 
\left(
\begin{array}{ccccccc}
\ddots & \vdots & \vdots & \vdots & \vdots & \vdots & \vdots \\
\ldots & O   & P_j & O   & O   & O   & \ldots \\
\ldots & Q_j & O   & P_j & O   & O   & \ldots \\
\ldots & O   & Q_j & O   & P_j & O   & \ldots \\
\ldots & O   & O   & Q_j & O   & P_j & \ldots \\
\ldots & O   & O   & O   & Q_j & O   & \ldots \\
\ldots & \vdots & \vdots & \vdots & \vdots & \vdots & \ddots
\end{array}
\right)
\qquad
\hbox{with}
\qquad
O = 
\left(
\begin{array}{cc}
0 & 0 \\
0 & 0 
\end{array}
\right),
\end{eqnarray*}
for $j = A$ and $B$. The unitary matrix $U_{(d)}$ for the discrete-time case corresponds to $U(t)$ for our continuous-time case at time $t=1$. More generally, $U_{(d)} ^n$ corresponds to $U(n)$ for $n=0,1, \ldots$. Once an explicit formula of $U(t)$ is obatined, the difference between continuous and discrete walks becomes clear. As we stated before, $U(t)$ has an infinite-range form. On the other hand, $U_{(d)}$ has a finite-range form. Moreover, we see that $U_{(d)}$ is not homogeneous. It is believed that the difference seems to be derived from the fact that discrete quantum walk has a coin but continuous quantum walk does not \cite{Amb}. However, the situation is not so simple, since the discrete-time case also does not necessarily need the coin (see Refs. \cite{HKS,Mey} for more detailed discussion).

We define a continuous-time quantum walk on $\ZM$ by $X_t$ whose probability distribution is defined by $P(X_t = k) = P(k,t)$ for any location $k \in \ZM $ and time $t \ge 0.$ Note that it follows from Corollary 2 that $P(k,t) = J_k ^2 (t)$. Then we obtain a new weak limit theorem for a contimuous-time quantum walk on the line: 

\begin{thm}
\label{main2}
If $-1 \le a < b \le 1$, then 
\begin{eqnarray*}
P ( a \le X_t/t \le b) \quad \to \quad \int_a ^b {1 \over \pi \sqrt{1 - x^2}} \> dx, \qquad (t \to \infty). 
\label{eqn:contilimit}
\end{eqnarray*}
\end{thm}
\noindent
Note that
\begin{eqnarray}
\int_{-1} ^1 {x^{2m} \over \pi \sqrt{1 - x^2}} \> dx
=
{2 \over \pi} \int_{0} ^{\pi/2} \sin ^{2m} \varphi \> d \varphi 
= {(2m-1)!! \over (2m)!!},
\label{eqn:moment}
\end{eqnarray}
for $m=1,2, \ldots ,$ where $n!!=n(n-2) \cdot \cdots \cdot 5 \cdot 3 \cdot 1,$ if $n=$ odd, $=n(n-2) \cdot \cdots \cdot 6 \cdot 4 \cdot 2,$ if $n=$ even. From Theorem 1 and Eq. (\ref{eqn:moment}), we have
\begin{cor}
For $m=1,2, \ldots ,$ 
\begin{eqnarray*}
E((X_t/t)^{2m}) \quad \to \quad (2m-1)!! / (2m)!!, \qquad (t \to \infty). 
\end{eqnarray*}
\end{cor}
By this corollary, for the standard deviation of our walk, $\sigma^q _{(c)} (t)$, we see that  
\begin{eqnarray*}
\sigma^q _{(c)} (t)/t \quad \to \quad 1/\sqrt{2} = 0.70710 \ldots, \qquad (t \to \infty). 
\end{eqnarray*}

We consider a {\it discrete-time} quantum walk $X^{(d)} _n$ with a symmetric distribution on the line, whose evolution is described by the Hadamard transformation (see Ref. \cite{NC}), that is,
\begin{eqnarray*}
U=
{1 \over \sqrt{2}}
\left(
\begin{array}{cc}
1 & 1 \\
1 & -1
\end{array}
\right).
\end{eqnarray*}
The walk is often called the Hadamard walk. In contrast with the continuous-time case, for the Hadamard walk, the following weak limit theorem holds \cite{GJS,Kon2002,Kon2005} : 
\begin{thm} 
\label{main2}
If $-1/\sqrt{2} \le a < b \le 1/\sqrt{2}$, then 
\begin{eqnarray*}
P ( a \le X^{(d)} _n /n \le b) \quad \to \quad \int_a ^b {1 \over \pi (1-x^2) \sqrt{1 - 2 x^2}} \> dx, \qquad (n \to \infty). 
\label{eqn:discretelimit}
\end{eqnarray*}
\end{thm}
As a corollary, we have 
\begin{eqnarray*}
\sigma^q _{(d)} (n)/n \quad \to \quad \sqrt{(2 - \sqrt{2})/2} = 0.54119 \ldots, \qquad (n \to \infty). 
\end{eqnarray*}
Comparing with the discrete-time case, the scaling in our continuous-time case is same, but the limit density function is slightly different. However, both density functions have some similar properties, for example, they have two peaks at the two end points of the support.

\section{Proof of Proposition 1}
To begin with, $A$ is rewritten as 
\begin{eqnarray}
A \> = \> 
\bordermatrix{
   & \ldots & -1 & \>\>\>0 & +1 & +2 & \ldots \cr
\vdots & \ddots & \vdots & \vdots & \vdots & \vdots & \vdots \cr
-1 & \ldots & T & P & O   & O   & \ldots \cr
\>\>\>0 & \ldots & Q & T & P & O   &  \ldots \cr
+1 & \ldots & O & Q & T & P & \ldots \cr
+2 & \ldots & O   & O   & Q & T &  \ldots \cr
\vdots & \ldots & \vdots & \vdots & \vdots & \vdots & \ddots
},
\label{eqn:kihon}
\end{eqnarray}
where 
\begin{eqnarray*}
P= 
\left(
\begin{array}{cc}
0 & 0 \\
1 & 0 
\end{array}
\right),
\quad 
T= 
\left(
\begin{array}{cc}
0 & 1 \\
1 & 0 
\end{array}
\right), 
\quad
Q=
\left(
\begin{array}{cc}
0 & 1 \\
0 & 0 
\end{array}
\right).
\end{eqnarray*}
This expression corresponds to a unitary matrix for a discrete-time quantum walk in which the particle maintains its position during each time step, (see \cite{HKS} for more details). The following algebraic relations are useful for some computations as in the case of a discrete-time walk:
\begin{eqnarray}
P^2 = Q^2 = O, \quad PT+TP=QT+TQ=I, \quad PQ+QP=T,
\label{eqn:benri}
\end{eqnarray}
where $O$ is $2 \times 2$ zero matrix and $I$ is $2 \times 2$ unit matrix. From now on, for simplicity, Eq. (\ref{eqn:kihon}) is written as 
\begin{eqnarray*}
A = [ \ldots, O, O, O, O, Q, T, P, O, O, O, O, O, \ldots ].
\label{eqn:kantan}
\end{eqnarray*}
A direct computation gives
\begin{eqnarray*}
&& 
A^2 = [ \ldots, O, O, O, O, I, 2I, I, O, O, O, O,  \ldots ], 
\\
&&
A^4 = [ \ldots, O, O, O, I, 4I, 6I, 4I, I, O, O, O, \ldots ],
\\
&&
A^6 = [ \ldots, O, O, I, 6I, 15I, 20I, 15I, 6I, I, O, O, \ldots ].
\label{eqn:kantan}
\end{eqnarray*}
It follows by induction that
\begin{eqnarray*}
A^{2n} = [ \ldots, O, O, A^{(2n)}_{-n} , \ldots, A^{(2n)}_{-1}, A^{(2n)}_{0}, A^{(2n)}_{1}, \ldots, A^{(2n)}_{n}, O, O,\ldots ],
\label{eqn:ippan}
\end{eqnarray*}
where $A^{(2n)}_{k}= A^{(2n)}_{-k} = a^{(2n)}_{k} I$ and 
\begin{eqnarray*}
a^{(2n)}_{k} = { 2n \choose n-k}, 
\label{eqn:ippankou}
\end{eqnarray*}
for any $k=0,1, \ldots, n.$ On the other hand, by using $A^{2n+1} = A^{2n} \times A$ and Eq. (\ref{eqn:benri}), we have
\begin{eqnarray*}
A^{2n+1} = [ \ldots, O, A^{(2n+1)}_{-(n+1)},  \ldots, A^{(2n+1)}_{-1}, A^{(2n+1)}_{0}, A^{(2n+1)}_{1}, \ldots, A^{(2n+1)}_{n+1}, O, \ldots ],
\label{eqn:ippan}
\end{eqnarray*}
where
\begin{eqnarray*}
&& 
A^{(2n+1)}_{-(n+1)} = a^{(2n)}_n Q, \quad 
A^{(2n+1)}_{-n} = a^{(2n)}_n T+ a^{(2n)}_{n-1} Q, 
\\
&&
A^{(2n+1)}_{-(n-1)} = (a^{(2n)}_{n-1}+a^{(2n)}_n )T+ (a^{(2n)}_{n-2}- a^{(2n)}_n ) Q, \ldots, 
\\
&&
A^{(2n+1)}_{-1} = (a^{(2n)}_{1}+a^{(2n)}_2 )T+ (a^{(2n)}_{0}- a^{(2n)}_2 ) Q, 
\quad A^{(2n+1)}_{0} = (a^{(2n)}_{0}+a^{(2n)}_1 )T, 
\\
&&
A^{(2n+1)}_{1} = (a^{(2n)}_{1}+a^{(2n)}_2 )T+ (a^{(2n)}_{0}- a^{(2n)}_2 ) P, \ldots, 
\\
&&
A^{(2n+1)}_{n} = a^{(2n)}_n T+ a^{(2n)}_{n-1} P, \quad A^{(2n+1)}_{(n+1)} = a^{(2n)}_n P.
\label{eqn:kisuu}
\end{eqnarray*}
The definition of $U(t)$ gives
\begin{eqnarray*}
U(t) 
&=& e^{itA/2} 
\> = \> \sum_{n=0}^{\infty} { \left( {it \over 2} \right)^{n} \over n! } A^n 
\\
&=& \sum_{n=0}^{\infty} (-1)^n { \left( {t \over 2} \right)^{2n} \over (2n)! } A^{2n} + i \sum_{n=0}^{\infty} (-1)^n { \left( {t \over 2} \right)^{2n+1} \over (2n+1)! } A^{2n+1}. 
\label{eqn:tenkai}
\end{eqnarray*}
Therefore we obtain $U(t) = B(t) + i C(t),$ 
where
\begin{eqnarray*}
&&
B(t) = [ \ldots , B_{-k}, \ldots, B_{-1}, B_{0}, B_{1}, \ldots, B_{k}, \ldots ],
\\
&&
C(t) = [ \ldots, C_{-k}, \ldots, C_{-1}, C_{0}, C_{1}, \ldots, C_{k}, \ldots ],
\label{eqn:cdayo}
\end{eqnarray*}
with
\begin{eqnarray*}
B_{k} = B_{-k} = b_k I, \qquad b_k = \sum_{n=0} ^{\infty} (-1)^n { \left( {t \over 2} \right)^{2n} \over (2n)! } {2n \choose n-k}, \quad (k=0,1, \ldots),
\label{eqn:bdayone}
\end{eqnarray*}
and 
\begin{eqnarray*}
&&
C_{-k} = c_k I + d_k Q, \qquad C_{k} = c_k I + d_k P, \quad (k=0,1, \ldots),
\\
&&
c_k = \sum_{n=0} ^{\infty} (-1)^n { \left( {t \over 2} \right)^{2n+1} 
\over (2n+1)! } \left\{ {2n \choose n-k} + {2n \choose n-(k+1)} \right\},
\quad (k=0,1, \ldots),
\\
&&
d_0 =0, 
\\
&&
d_k = \sum_{n=0} ^{\infty} (-1)^n { \left( {t \over 2} \right)^{2n+1} 
\over (2n+1)! } \left\{ {2n \choose n-(k-1)} + {2n \choose n-(k+1)} \right\},
\quad (k=1,2, \ldots).
\label{eqn:cdayone}
\end{eqnarray*}
Remark that
\begin{eqnarray}
J_k (x) = \left( {x \over 2} \right)^k \sum_{m=0} ^{\infty} { \left( {ix \over 2} \right)^{2m} \over (k+2m)! } {k+2m \choose m}, 
\label{eqn:jyuyou}
\end{eqnarray}
(see Eq. (4.9.5) in Ref. \cite{AAR}). Finally, by using Eq. (\ref{eqn:jyuyou}) and the following relation: 
\begin{eqnarray*}
{ 2n \choose k } + { 2n \choose k-1 } = { 2n+1 \choose k },
\end{eqnarray*}
we have the desired conclusion.

\section{Proof of Theorem 1}
We begin by stating the following result (see page 214 in Ref. \cite{AAR}): suppose that $a,b,$ and $c$ are lengths of sides of a triangle and $c^2 = a^2+b^2 - 2ab \cos \xi.$ Then 
\begin{eqnarray}
J_0 (c) = \sum_{k= -\infty} ^{\infty} J_k (a) J_k (b) e^{ik \xi}.
\label{eqn:triangledayo}
\end{eqnarray}
If we set $t=a=b$ in Eq. (\ref{eqn:triangledayo}), then 
\begin{eqnarray}
J_0 (t \> \sqrt{2(1-\cos \xi)}) = \sum_{k= -\infty} ^{\infty} J_k ^2 (t) e^{i k \xi}.
\label{eqn:triangledazo}
\end{eqnarray}
From Eq. (\ref{eqn:triangledazo}), we see that the characteristic function of a continuous-time quantum walk on the line is given by
\begin{eqnarray}
E(e^{i \xi X_t}) =  \sum_{k= -\infty} ^{\infty}  e^{i k \xi} J_k ^2 (t) = 
J_0 (t \> \sqrt{2(1-\cos \xi)}).
\label{eqn:triangledane}
\end{eqnarray}
First we consider that $t$ is a positive integer case, that is, $n(=t)=1,2,\ldots.$ By using Eq. (\ref{eqn:triangledane}), we have
\begin{eqnarray*}
E(e^{i \xi X_n/n}) = J_0 (n \> \sqrt{2(1-\cos (\xi/n))}) \to J_0 (\xi), 
\label{eqn:triangledanedane}
\end{eqnarray*}
as $n \to \infty.$ To know a limit density funtion, we use the following expression of $J_0 (x)$ (see Eq. (4.9.11) in Ref. \cite{AAR});   
\begin{eqnarray}
J_0 (\xi) = {1 \over \pi} \int_0 ^{\pi} \cos (\xi \sin \varphi) \> d \varphi = {2 \over \pi} \int_0 ^{\pi/2} \cos (\xi \sin \varphi) \> d \varphi.
\label{eqn:iine}
\end{eqnarray}
Taking $x= \sin \varphi$, we have
\begin{eqnarray}
J_0 (\xi) = \int_{-1} ^1 \cos (\xi x) { 1 \over \pi \sqrt{1 - x^2} } \> dx 
= \int_{-1} ^1 e^{i \xi x} { 1 \over \pi \sqrt{1 - x^2} } \> dx. 
\label{eqn:iiyone}
\end{eqnarray}
By using Eq. (\ref{eqn:iine}), we get
\begin{eqnarray*}
|J_0 (\xi) - J_0 (0)| \le {2 \over \pi} \int_0 ^{\pi/2} | \cos (\xi \sin \varphi) -1| \> d \varphi.
\label{eqn:iinene}
\end{eqnarray*}
From the bounded convergence theorem, we see that the limit $J_0 (\xi)$ is continuous at $\xi = 0$, since $\cos (\xi \sin \varphi) \to 1 $ as $\xi \to 0$. Therefore, by the continuity theorem (see page 99 in \cite{Dur}) and Eq. (\ref{eqn:iiyone}), we conclude that if $n \to \infty$, then $X_n / n$ converges weakly to a random variable whose density function is given by $1 /\pi \sqrt{1 - x^2}$ for $x \in (-1, 1)$. That is, if $-1 \le a < b \le 1$, then 
\begin{eqnarray*}
P ( a \le X_n /n \le b) \quad \to \quad \int_a ^b {1 \over \pi \sqrt{1 - x^2}} \> dx,
\end{eqnarray*}
as $n \to \infty.$ Next, to deal with values of $t$ that are not integers, we want to show
\begin{eqnarray*}
\left| P ( a \le X_t /t \le b) - P ( a \le X_{[t]} / [t] \le b) \right| \quad \to \quad 0, \qquad (t \to \infty),
\end{eqnarray*}
where $[x]$ denotes the integer part of $x$. To do this, we observe that
\begin{eqnarray*}
&&
\left| P ( a \le X_t /t \le b) - P ( a \le X_{[t]} / [t] \le b) \right| 
\\
&&
= 
\left| P ( a t \le X_t \le bt ) - P ( a [t] \le X_{[t]} \le b [t] ) \right| 
\\
&&
=
\left| \sum_{at \le k \le bt} P (X_t = k) - \sum_{a[t] \le k \le b[t]} P ( X_{[t]}=k) \right| 
\\
&&
=
\left| \sum_{at \le k \le bt} J_k ^2 (t) - \sum_{a[t] \le k \le b[t]} J_k ^2 ([t]) \right| 
\\
&&
\le
\left| \left( \sum_{at \le k \le bt} - \sum_{a[t] \le k \le b[t]} \right) J_k ^2 (t)  \right|
+ 
\left| \sum_{a[t] \le k \le b[t]} ( J_k ^2 (t) - J_k ^2 ([t]) ) \right|
\\
&&
= K_1 (t) + K_2 (t). 
\end{eqnarray*}
To estimate $K_1 (t)$ and $K_2 (t)$, we use the following Meissel's second expansion (see Eq. (5) in page 228 of \cite{Wat}):
\begin{eqnarray}
J _{\nu} (\nu \sec \beta) = \sqrt{{2 \cot \beta \over \pi \nu}} \> 
e^{- P(\nu, \beta)} \> \cos (Q(\nu,\beta) -\pi/4),
\label{eqn:meissel}
\end{eqnarray}
where $P(\nu, \beta)$ and $Q(\nu, \beta)$ are defined by Eqs. (1) and (2) respectively in page 228 of \cite{Wat}.

First we consider $K_1(t)$ case. We see that
\begin{eqnarray}
K_1 (t) 
&\le& 
\max \{J^2 _{[at]}(t), J^2 _{[bt]}(t), |J^2 _{[bt]}(t)-J^2 _{[at]}(t)| \}
\nonumber
\\
&\le& 
2 \times \max \{J^2 _{[at]}(t), J^2 _{[bt]}(t) \}.
\label{eqn:kin}
\end{eqnarray}
Without a loss of generality, we consider only $J^2 _{[ct]}(t)$ with $0 < c <1.$ From Eq. (\ref{eqn:meissel}), taking $\nu = [ct]$, we have
\begin{eqnarray}
J^2 _{[ct]} (t) = J^2 _{[ct]} ([ct] \sec \beta) = {2 \cot \beta \over \pi [ct]} \> e^{- 2 P([ct], \beta)} \> \cos^2(Q([ct],\beta) -\pi/4),
\label{eqn:meisseltwo}
\end{eqnarray}
where $\sec \beta = t/[ct] (>1).$ Then, for large $t$, we get
\begin{eqnarray*}
J^2 _{[ct]} (t) \le {2 \over \pi [ct]} \times {2c \over \sqrt{1 - c^2}},
\label{eqn:meisselthree}
\end{eqnarray*}
since $\cot \beta \to c/\sqrt{1-c^2}$ as $t \to \infty.$ The last inequality implies
\begin{eqnarray}
J^2 _{[ct]} (t) \to 0, \qquad (t \to \infty).
\label{eqn:meisselthree}
\end{eqnarray}
Combining Eq. (\ref{eqn:kin}) with Eq. (\ref{eqn:meisselthree}) implies that $K_1(t) \to 0$ as $t \to \infty.$

Next we consider $K_2 (t)$ case. Noting Eq. (\ref{eqn:meisselthree}), it is enough to show that
\begin{eqnarray*}
\left| \sum_{[at] \le k \le [bt]}  ( J_k ^2 (t) -  J_k ^2 ([t]) ) \right| \to 0, \qquad (t \to \infty).
\end{eqnarray*}
As in the case of $K_1 (t)$, fix $\> 0 <c <1$. Define $\beta_1$ and $\beta_2$ by $\sec \beta_1 =t/[ct]$ and  $\sec \beta_2 =[t]/[ct]$ respectively. Then we have\begin{eqnarray*}
&&
\cot \beta_1 \sim \cot \beta_2 \sim {c \over \sqrt{1-c^2}},
\\
&&
\cot \beta_1 - \cot \beta_2 
\sim -{c \over (1-c^2)^{3/2}} \times {t-[t] \over t},
\\
&&
\tan \beta_1 \sim \tan \beta_2 \sim {\sqrt{1-c^2} \over c},
\\
&&
\tan \beta_1 - \tan \beta_2 
\sim {1 \over c \sqrt{1-c^2}} \times {t-[t] \over t},
\\
&&
\beta_1 \sim \beta_2 \sim \arccos c,
\\
&&
\beta_1 - \beta_2 
\sim {(2 - \sqrt{1-c^2}) c \over 2 \sqrt{1-c^2}} \times {t-[t] \over t},
\label{eqn:ganbare}
\end{eqnarray*}
where $f(t) \sim g(t)$ means $f(t)/g(t) \to 1$ as $t \to \infty$. By using the above estimates, Eq. (\ref{eqn:meisseltwo}), and noting $P([ct], \beta_i) \to 0 \>(t \to \infty)$, $\>Q([ct], \beta_i) \sim [ct] (\tan \beta_i - \beta_i)$ for $i=1,2$, we see that
\begin{eqnarray*}
&& 
J^2 _{[ct]} (t) - J^2 _{[ct]} ([t]) =
J^2 _{[ct]} ([ct] \sec \beta_1) - J^2 _{[ct]} ([ct] \sec \beta_2)
\\
&\sim& 
{2 \over \pi [ct]} \{ \cot \beta_1 \cos^2 ([ct] (\tan \beta_1 - \beta_1)-\pi/4)
\\
&&
\qquad \qquad \qquad \qquad 
- \cot \beta_2 \cos^2 ([ct] (\tan \beta_2 - \beta_2)-\pi/4) \}
\\
&\sim& 
{1 \over \pi [ct]} \{ \cot \beta_1 - \cot \beta_2 
\\
&&
\qquad \quad 
+ \cot \beta_1 \sin (2 [ct] (\tan \beta_1 - \beta_1)) - \cot \beta_2 \sin (2 [ct] (\tan \beta_2 - \beta_2)) \}\\
&\sim& 
{1 \over \pi [ct]} [ (\cot \beta_1 - \cot \beta_2) \{ 1+ \sin (2 [ct] (\tan \beta_1 - \beta_1)) \} 
\\
&&
\qquad \quad 
+ \cot \beta_2 \{ \sin (2 [ct] (\tan \beta_1 - \beta_1)) - \sin (2 [ct] (\tan \beta_2 - \beta_2)) \} ]
\\
&\sim& 
{1 \over \pi [ct]} [ (\cot \beta_1 - \cot \beta_2) \{ 1+ \sin (2 [ct] (\tan \beta_1 - \beta_1)) \}
\\
&&
\qquad \quad 
- \cot \beta_2 \times 2 [ct] \{ (\tan \beta_1 - \tan \beta_2) - (\beta_1 - \beta_2) \} \cos (2 [ct] (\tan \beta_1 - \beta_1)) ].
\end{eqnarray*}
Therefore we obtain
\begin{eqnarray}
&&
J^2 _{[ct]} (t) - J^2 _{[ct]} ([t]) 
\nonumber
\\ 
&&
\quad
\sim \> - 
{t - [t] \over \pi (1-c^2) t^2} 
\left\{ C_1 (c) (1 + \sin (\phi (c,t)) ) + t \> C_2 (c) \cos (\phi (c,t)) \right \},
\label{eqn:ganbare}
\end{eqnarray}
where $C_1 (c) = 1/\sqrt{1-c^2},$ $\> C_2 (c) = 2 - (2 - \sqrt{1-c^2})c^2,$ and $\> \phi (c,t) = 2 [ct]$ $( \sqrt{1-c^2}/c $ $ - \arccos c).$ For simplicity, we suppose that $0 < a < b \le 1$. Noting Eq. (\ref{eqn:meisselthree}), we see that
\begin{eqnarray*}
\lim_{t \to \infty} K_2 (t)
&=&
\lim_{t \to \infty} \left| \sum_{k=0} ^{[bt]-[at]} 
\left( J^2 _{[at]+k} (t) -  J^2 _{[at]+k} ([t]) \right) \right|
\\
&=&
\lim_{t \to \infty} (b-a)t \left| \int_{0} ^{1} 
\left( J^2 _{[(a+(b-a)x)t]} (t) -  J^2 _{[(a+(b-a)x)t]} ([t]) \right) \> dx \right|\\
&\le&
\lim_{t \to \infty} {b-a \over \pi t} 
\int_{0} ^{1} \left| { C_1 (a+(b-a)x) \{1 + \sin (\phi (a+(b-a)x,t)) \} \over 1- (a+(b-a)x)^2} \right| \> dx 
\\
&+&
\lim_{t \to \infty} {b-a \over \pi} 
\left| \int_{0} ^{1} { C_2 (a+(b-a)x) \over 1- (a+(b-a)x)^2} \> \cos (\phi (a+(b-a)x,t)) \> dx \right|
\\
&=&
\lim_{t \to \infty}  K_3 (t) + \lim_{t \to \infty} K_4 (t).
\end{eqnarray*}
To get the third inequality, we used Eq. (\ref{eqn:ganbare}). It is easily obtained that $\lim_{t \to \infty}$ $\> K_3 (t) =0,$ since the integral is bounded above by a constant which is independent of $t$. On the other hand, it follows from the Riemann-Lebesgue lemma that $\lim_{t \to \infty} K_4 (t) = 0.$ So we have $\lim_{t \to \infty} K_2 (t) =0.$ Thus the proof of Theorem 1 is complete.

\section{Conclusion and Discussion}
In contrast with the classical random walk for which the central limit theorem holds, we have shown a weak limit theorem $X^{(c)}_t /t  \to dx / \pi \sqrt{1 - x^2} \> (t \to \infty)$ for the continuous-time quantum walk on the line (Theorem 1). Interestingly, although the definition of the walk is very different from that of discrete-time one, the limit theorem resembles that of discrete-time case, for example, the symmetric Hadamard walk: $X^{(d)} _n /n  \to dx / \pi (1-x^2) \sqrt{1 - 2 x^2} \> (n \to \infty).$ 

Very recently, Romanelli {\it et al.} \cite{RSSAAD} investigated a continuum time limit for a discrete-time quantum walk on $\ZM$ and obtained the position probability distribution. When the initial condition is given by $\tilde{a}_l (0) = \delta_{l,0}, \tilde{b}_l (0) \equiv 0$ in their notation for the Hadamard walk, the distribution at location $k$ and time $t$ becomes the following in our notation: $P_{(R)} (k,t) = J_{k}^2(t/\sqrt{2}).$ More generally, we consider the time evolution given by the following unitary matrix:
\begin{eqnarray*}
U (\theta) =
\left(
\begin{array}{cc}
\cos \theta & \sin \theta \\
\sin \theta & - \cos \theta
\end{array}
\right),
\end{eqnarray*}
\par\noindent
where $\theta \in (0, \pi/2).$ Note that $\theta = \pi/4$ case is equivalent to the Hadamard walk. Then we have $P_{(R, \theta)} (k,t) = J_{k}^2(t \> \cos \theta).$ 
In this case, a similar argument in the proof of Theorem 1 implies that if $- \cos \theta \le a < b \le \cos \theta$, then 
\begin{eqnarray*}
P ( a \le X^{(R,\theta)} _t/t \le b) \quad \to \quad \int_a ^b {1 \over \pi \sqrt{\cos^2 \theta- x^2}} \> dx, \qquad (t \to \infty), 
\label{eqn:contilimit}
\end{eqnarray*}
where $X^{(R,\theta)} _t$ denotes a continuous-time quantum walk whose probability distribution is given by $P_{(R,\theta)} (k,t).$ As a consequence, we obtain
\begin{eqnarray*}
E((X^{(R,\theta)} _t/t)^{2m}) \quad \to \quad \cos^{2m} \theta \times (2m-1)!! / (2m)!!, \qquad (t \to \infty). 
\end{eqnarray*}
In particular, when $m=1$, the limit $\cos^{2} \theta /2$ is consistent with Eq. (30) in Romanelli {\it et al.} \cite{RSSAAD}


\begin{thebibliography}{99}

\bibitem{Amb} A. Ambainis, International Journal of Quantum Information, {\bf 1}, 507 (2003).

\bibitem{Kem} J. Kempe, Contemporary Physics, {\bf 44}, 307 (2003).

\bibitem{TFMK} B. Tregenna, W. Flanagan, W. Maile and V. Kendon, New Journal of Physics, {\bf 5}, 83 (2003).

\bibitem{AAKV} D. Aharonov, A. Ambainis, J. Kempe, and U. V. Vazirani, Proceedings of the 33rd Annual ACM Symposium on Theory of Computing, 50 (2001).

\bibitem{ABNVW} A. Ambainis, E. Bach, A. Nayak, A. Vishwanath, and J. Watrous, Proceedings of the 33rd Annual ACM Symposium on Theory of Computing, 37 (2001).

\bibitem{GJS} G. Grimmett, S. Janson, and P. F. Scudo, Phys. Rev. E, \textbf{69}, 026119 (2004).

\bibitem{HKS} M. Hamada, N. Konno, and E. Segawa, RIMS Kokyuroku, No.1422, 1 (2005), e-print quant-ph/0408100.

\bibitem{Kon2002} N. Konno, Quantum Information Processing, \textbf{1}, 345 (2002).

\bibitem{Kon2005} N. Konno, J. Math. Soc. Jpn., \textbf{57}, 1179 (2005), e-print quant-ph/0206103.

\bibitem{Mey} D. A. Meyer, J. Statist. Phys., \textbf{85}, 551 (1996).

\bibitem{PRR} A. Patel, K. S. Raghunathan, and P. Rungta, e-print quant-ph/0405128.

\bibitem{RSSAAD} A. Romanelli, A. C. Sicardi Schifino, R. Siri, G. Abal, A. Auyuanet, and R. Donangelo, Physica A, \textbf{338}, 395 (2004).

\bibitem{AAHT} W. Adamczak, K. Andrew, P. Hernberg, and C. Tamon, e-print quant-ph/0308073.

\bibitem{ABTW} A. Ahmadi, R. Belk, C. Tamon, and C. Wendler, Quantum Information and Computation, \textbf{3}, 611 (2003).

\bibitem{CCDFGS}  A. M. Childs, R. Cleve, E. Deotto, E. Farhi, S. Gutmann, and D. A. Spielman, Proceedings of the 35th Annual ACM Symposium on Theory of Computing, 59 (2003).

\bibitem{CFG} A. M. Childs, E. Farhi, and S. Gutmann, Quantum Information Processing, \textbf{1}, 35 (2002). 

\bibitem{FG} E. Farhi and S. Gutmann, Phys. Rev. A, \textbf{58}, 915 (1998).


\bibitem{IKKK} N. Inui, K. Kasahara, Y. Konishi and N. Konno, Fluctuation and Noise Letters, \textbf{5}, 73 (2005).

\bibitem{MR} C. Moore and A. Russell, e-print quant-ph/0104137.

\bibitem{Dur} R. Durrett, \textit{Probability: Theory and Examples, 2nd edition} (Duxbury Press, 1996).

\bibitem{GS} G. R. Grimmett and D. R. Stirzaker, \textit{Probability and Random Processes, 2nd edition} (Oxford University Press, 1992).

\bibitem{GW} H. Gerhardt and J. Watrous, e-print quant-ph/0305182, to appear at RANDOM'03.

\bibitem{Wat} G. N. Watson, \textit{A Treatise on the Theory of Bessel Functions, 2nd edition} (Cambridge University Press, Cambridge, 1944).

\bibitem{AAR} G. E. Andrews, R. Askey and R. Roy, \textit{Special Functions} (Cambridge University Press, 1999).

\bibitem{NC} M. A. Nielsen and I. L. Chuang, \textit{Qunatum Computation and Quantum Information} (Cambridge University Press, 2000).

\end{thebibliography}
\end{document}